\documentclass[a4paper]{VisionStyle}
\usepackage{epsfig}

\begin{document}

\title{Timing analysis of the isolated neutron star
\object{RX~J0720.4-3125}}

\author{S.\,Zane\inst{1}, F.\, Haberl\inst{2}, M.\, Cropper\inst{1}, \,
V. Zavlin\inst{2}, D.\,
Lumb\inst{3}, \, S. \, Sembay\inst{4},  \and C.\,Motch\inst{5}} 

\institute{
Mullard Space Science Laboratory, University College
London,  Holmbury St. Mary, Dorking, Surrey, RH5 6NT, UK
\and                                                  
Max Planck Institut f\"ur Extraterrestrische Physik,
Giessenbachstrasse, D-85748 Garching, Germany
\and 
Space Science Department, ESTEC, Posbus 299, Keplerlaan 1, Noordwijk
2200 AG, The Netherlands
\and 
X-ray Astronomy Group, Department of Physics and Astronomy, University
of Leicester, Leicester LE1 7RH, UK 
\and
Observatoire Astronomique, CNRS UMR 7550, 11 Rue de l'Universit\'e,
F-67000 Strasbourg, France}

\maketitle 

\begin{abstract}

We present a combined analysis of {\sl XMM-Newton}, {\sl Chandra} and {\sl
Rosat} observations of the isolated neutron star \object{RX~J0720.4-3125},
spanning
a total period of $\sim 7$ years. We develop a maximum likelihood
periodogramme based on $\Delta
C$-statistic and maximum likelihood method, which are appropriate
for sparse event lists. As an {\it a posteriori} check, we have folded a
further {\sl BeppoSAX}-dataset with the period predicted at the time of
that observation, finding that the phase is consistent.       

The study of the spin history and the measure of the spin-down rate are of
extreme importance in understanding the mechanism powering
\object{RX~J0720.4-3125}. The value of $\dot P$,
here measured for the first time, is $\approx 10^{-14}$~s/s and can not be  
explained in terms of propeller or torque from a fossil disk. When
interpreted in terms of dipolar losses, it gives a magnetic field of $B
\approx 10^{13}$~G,
making also implausible that the source is accreting from the underdense
surroundings.
We also find unlikely that the field decayed
from a much larger value ($B\approx 10^{15}$~G in the past, as expected in
the ``old magnetar'' interpretation) since this
scenario predicts a source age of $\approx 10^4$~yrs, too young to match
the (low) observed X-ray luminosity. The observed properties are more
compatible with a scenario in which the source is $\approx
10^6$~yrs old, and its magnetic field has not changed substantially
over the lifetime.

\end{abstract}

\section{Introduction}

\object{RX~J0720.4-3125} is a nearby, isolated neutron star (NS) detected
by {\sl
ROSAT} during a Galactic plane survey (\cite{szane-D1_2:ha97}) 
and recently
re-observed with {\sl XMM-Newton} on 2000 May 13 (\cite{szane-D1_2:pa01}, 
\cite{szane-D1_2:cr01}) and 2001 November 21. The source exhibits all the
common characteristics of the other six ROSAT NS candidates (hereafter dim
NSs, see \cite{szane-D1_2:tr00} for a review): a blackbody-like spectrum
with $kT\sim 80$~eV; a large X-ray to optical flux
ratio; a low X-ray luminosity, $L_{\rm X}\approx 10^{30}-10^{31}$~erg/s; 
a low column density and no evidence for a binary
companion. In addition, \object{RX~J0720.4-3125} is pulsating with a
period $P \sim
8.4$~s. 

Until a few years ago, dim NSs were thought to
constitute a class stand alone and two mechanisms were proposed for their
emission: either accretion from the interstellar medium onto an old NS or
release of thermal radiation from a younger, cooling
object. More recently, based on the similarity of the periods, it has been
suggested a possible evolutionary link between dim
NSs, anomalous X-ray pulsars (AXPs), and soft gamma-ray repeaters (SGRs).
Two kind of ``unified''
scenarios have been then proposed. In the first one, 
the three classes of objects are powered by dissipation of a decaying,
superstrong magnetic field ($B\geq 10^{14}-10^{15}$~G). In this case
dim NSs are the descendants of SGRs and AXPs, and \object{RX~J0720.4-3125}
may be one of the closest old-magnetars.
Alternatively, all the three classes may contain NSs with lower
(canonical) magnetic field ($B \approx
10^{12}~G$) endowed by a fossil disk (\cite{szane-D1_2:aay01}). In
this case dim NSs in the propellor phase would be the
progenitors of AXPs and SGRs, the latter having entered an accretion
phase.                                                

Recently, \cite*{szane-D1_2:pa01} presented {\sl XMM-Newton} spectra of
\object{RX~J0720.4-3125}. The absence of electron or proton
cyclotron resonances in the RGS range excluded 
magnetic fields of $B \approx (0.3-2)\times
10^{11}$~G and $(0.5-2)\times 10^{14}$~G (see \cite{szane-D1_2:za01}).
Based on
the same {\sl XMM-Newton} observation, \cite*{szane-D1_2:cr01} presented
the
pulse-shape analysis. They derived an upper limit
on the polar cap size,  showing that an emitting region larger
than $\sim 60\degr-65\degr$ can be rejected at a confidence level of 90\%. 
Whatever the
mechanism, the X-ray emitting region is therefore confined to a
relatively small fraction of the star surface.
They also found that the 
hardness ratio is softest around the flux maximum. The same has been later
discovered by \cite*{szane-D1_2:pe01} in some AXPs. 
\cite*{szane-D1_2:cr01} suggested two possible
explanations for this effect: either radiation beaming (as in their
best-fitting model) or the presence of a spatially variable
absorbing matter, co-rotating in the magnetosphere. The latter may be
indeed the case if the star is propelling matter outward 
(\cite{szane-D1_2:aay01}).  

Further information about the nature of this puzzling source can be
obtained by the spin history. Magnetars will spin-down at a rate 
$\dot P \approx 10^{-11}(B/10^{14}\, {\rm
G})^2/P$ ss$^{-1}$, due to 
magneto-dipolar losses. The preliminary measure of $\dot P$
published by \cite*{szane-D1_2:ha97} for \object{RX~J0720.4-3125} is
uncertain to a
considerable large value, and does not even allow spin-up and spin-down
to be discriminated. An accurate determination of $\dot P$ is
therefore crucial, as well a tracking of the spin history of the source.
Here we present a combined analysis of {\sl XMM-Newton}, {\sl
Chandra} and {\sl Rosat} data, spanning a period of $\sim 7$ years. For
all details we refer to the paper \cite*{szane-D1_2:za02}. 

\section{Timing analysis}
\label{szane-D1_2_sec:obs}

\begin{figure}[ht]
  \centerline{
    \epsfig{file=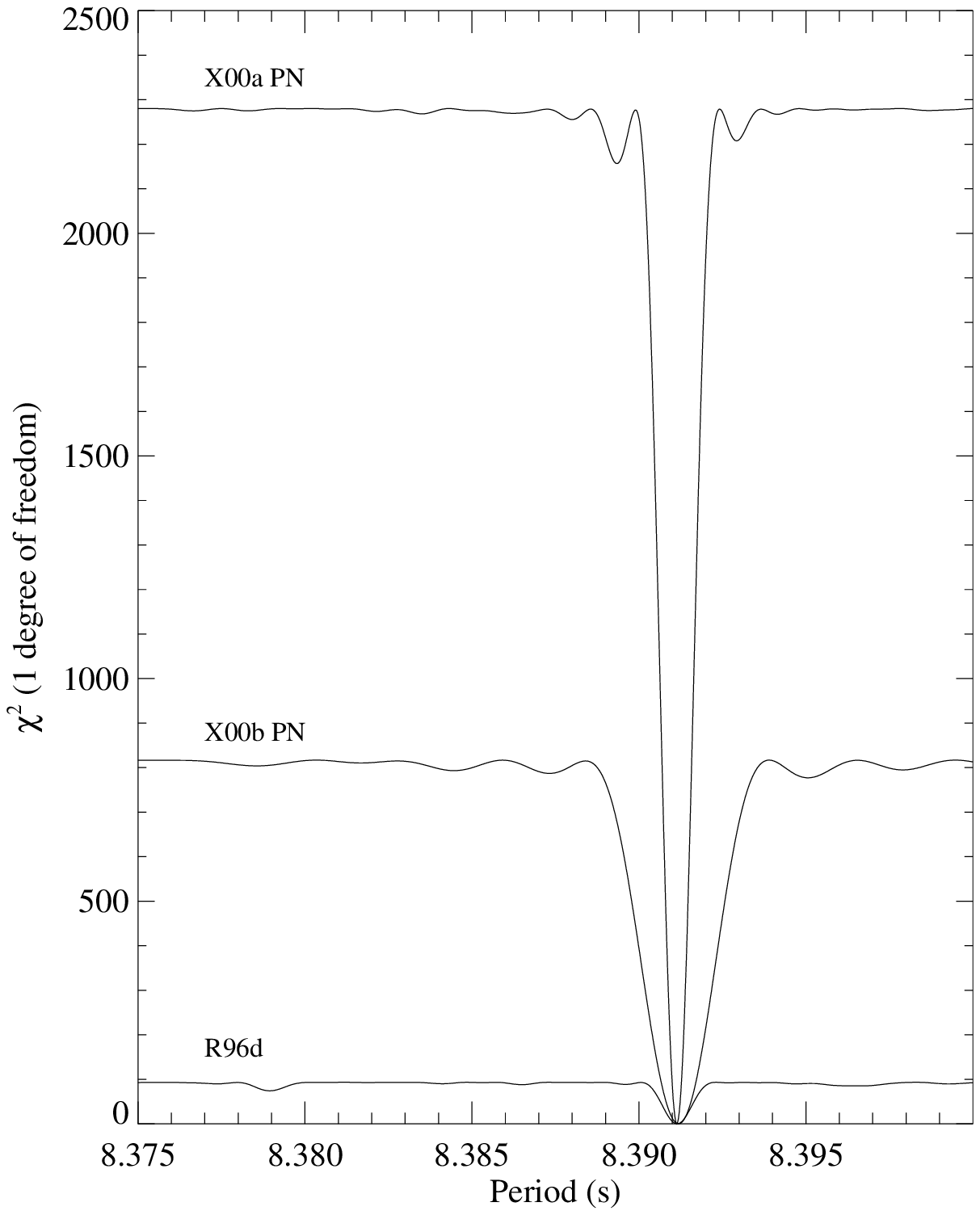, width=4.5cm}
    \epsfig{file=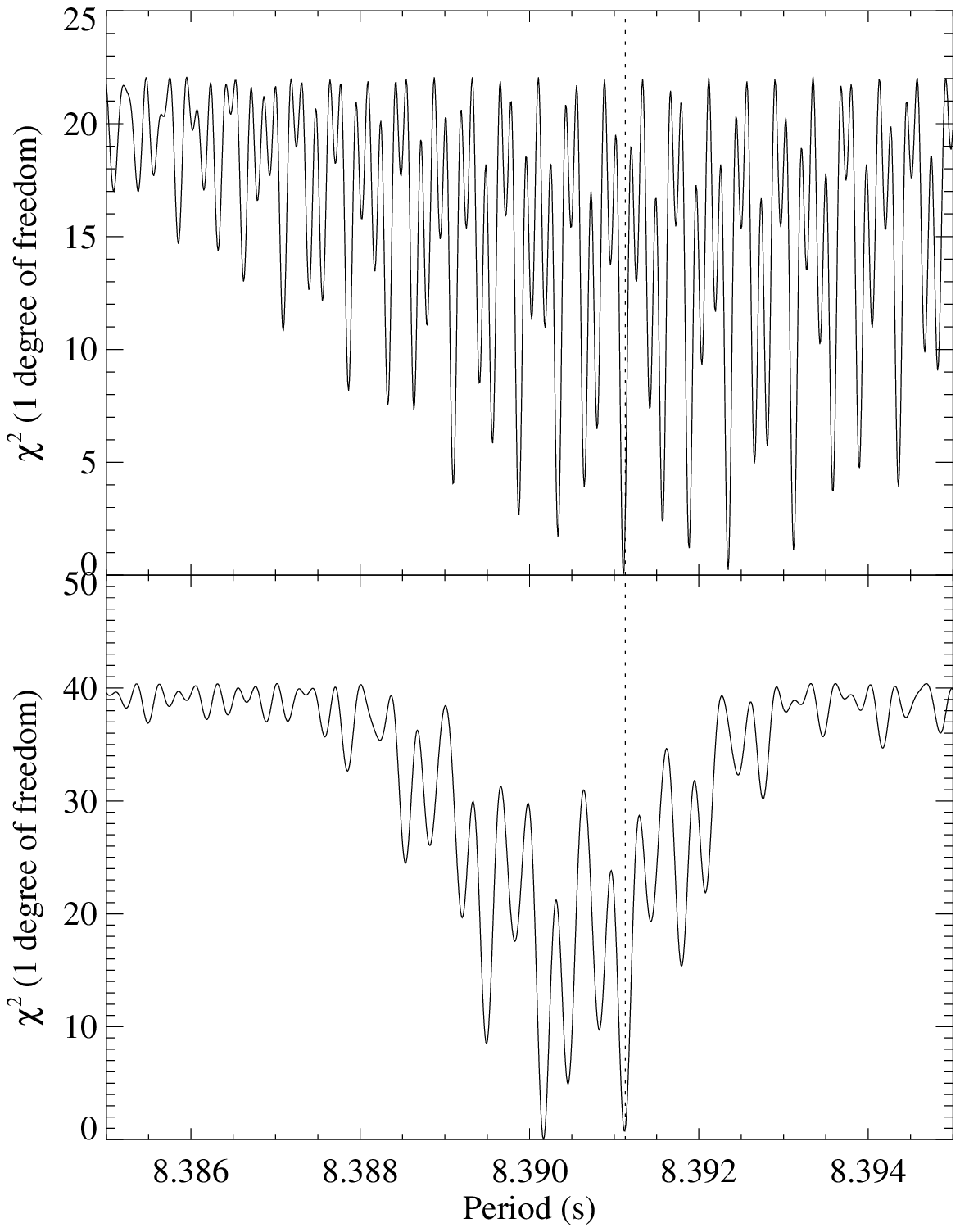, width=4.5cm}}
\caption{ 
(Left) Maximum likelihood periodogrammes (MLP) for three long datasets,
R96d, X00a (PN) and X00b (PN), showing the 
periodicity at 8.391~s. These constrain the selection of the strongest and
second-strongest dips in the MLPs for the R98 and Ch00 datasets
respectively (right). The vertical line denotes a period of 8.39113~s. The
68\% and 90\% confidence levels are at $\chi ^2 = 1.0$ and 2.71 for one
degree of freedom.}  
\label{szane-D1_2_fig:fig1}
\end{figure}

The different observations used in our analysis are shown in table
\ref{szane-D1_2_tab:tab1}; the major datasets are from the two {\sl XMM}
observations and from the 1996~Nov.~3 {\sl Rosat} pointing, while the 1998
{\sl Rosat} and the {\sl Chandra} observations are valuable by nature of
their several day durations. Our data
originate from instrumentation with widely differing sensitivities: 
typical count rates vary from 1 count every $\sim 3$~s for {\sl Rosat} HRI
to $\sim 6$ counts/s for {\sl XMM-Newton} PN. However, none of these count
rates is sufficiently high for a normal distribution of counts to be
expected, thus standard discrete Fourier Transforms are not directly 
applicable. For sparse data and event list data, we used instead 
Rayleigh Transform (i.e. \cite{szane-D1_2:dj91}, \cite{szane-D1_2:ma72}). 
It is also crucial for us to define precisely the confidence
intervals to the derived quantities, in particular the period $P$. We do
this by constructing MLP (maximum likelihood periodogrammes) which make no
assumptions on data distribution, and using the $\Delta C$-statistics
(\cite{szane-D1_2:ca79}). The uncertainty in the period and the $\chi ^2$
can be read directly from the $y$-axis of the MLP (see figure
\ref{szane-D1_2_fig:fig1}). 

\begin{table}[bht]
  \caption{The {\sl ROSAT}, {\sl Chandra} and {\sl XMM-Newton}
observations of \object{RX~J0720.4-3125} used in this analysis. The entry
in the
third column is the effective exposure.} 
  \label{szane-D1_2_tab:tab1}
  \begin{center}
    \leavevmode
    \footnotesize
    \begin{tabular}[h]{llcc}
      \hline \\[-5pt]
Date & Instrument & Eff. Exp. (s) &
Label \\ [+5pt]
      \hline \\[-5pt]
1993 Sep 27 & {\sl Rosat} PSPC &  3221 & R93\\
1996 Apr 25 & {\sl Rosat} HRI &  3566 & R96a \\
1996 May 7  & {\sl Rosat} HRI &  3125 &
R96b \\
1996 Sep 27 & {\sl Rosat} HRI  &  1409 &
R96c \\
1996 Nov 3  & {\sl Rosat} HRI  & 33569 &
R96d \\
1998 Sep 27 & {\sl Rosat} HRI &  3566 & R98
\\
2000 Feb 1  & {\sl Chandra} HRC-S &
  37635 & Ch00 \\
2000 May 13 & {\sl XMM} MOS1 &
  61648 & X00a \\
            &       ~~~~~~~~~MOS2 & 61648
\\
            &       ~~~~~~~~~PN & 62425
\\ 
2000 Nov 21 & {\sl XMM} MOS1 
& 17997 & X00b \\
            &       ~~~~~~~~~MOS2 &
17994 \\
            &       ~~~~~~~~~PN & 25651
\\
\hline
\\
1997 Mar 16 & {\sl SAX} LECS &    & S97
\\
      \hline \\
      \end{tabular}
  \end{center}
\end{table}

\begin{figure}[bht]
\begin{center}
\setlength{\unitlength}{1cm}
\begin{picture}(11,6)
\put(-1.,-1.5){\includegraphics{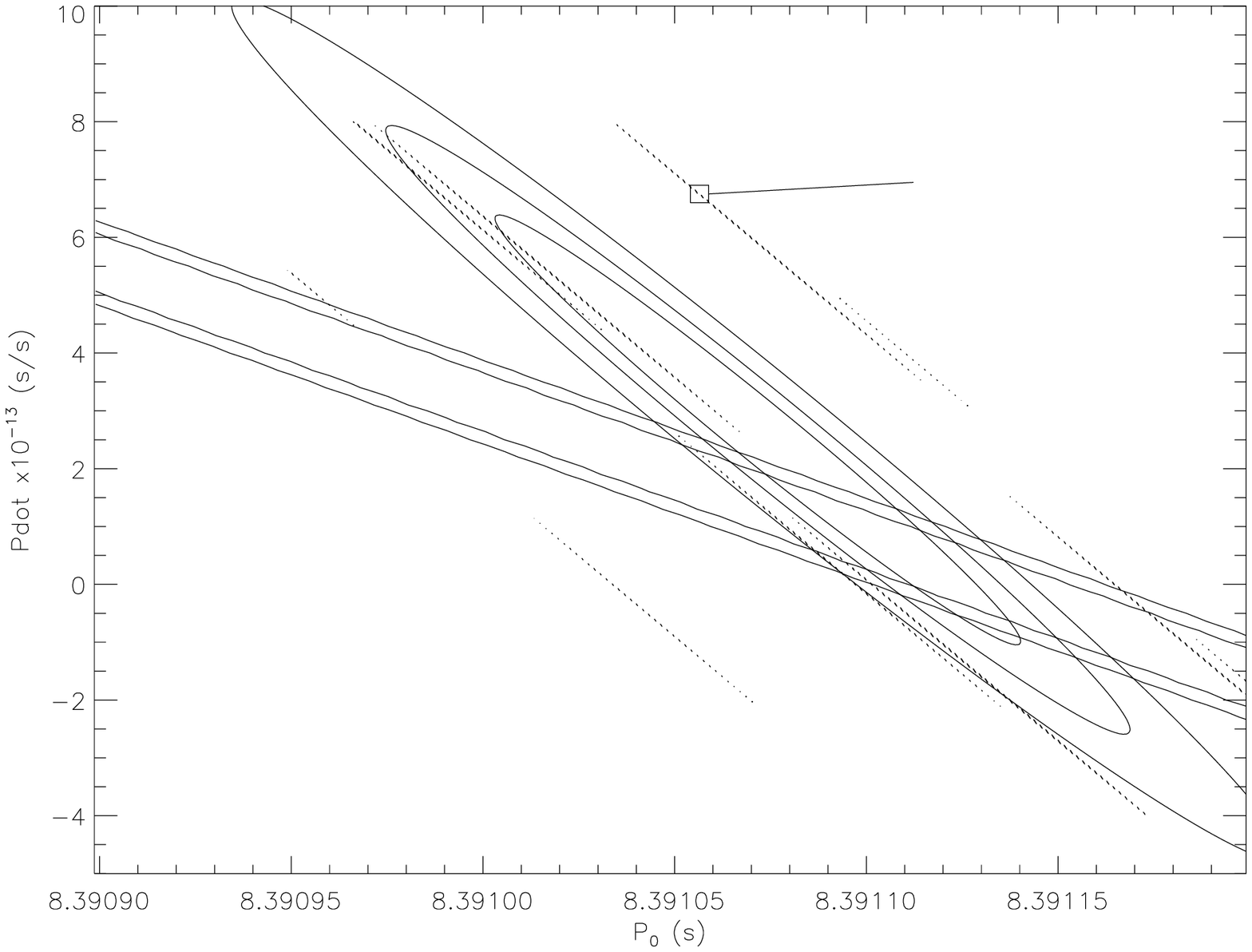}}
\put(6.,3.3){\includegraphics{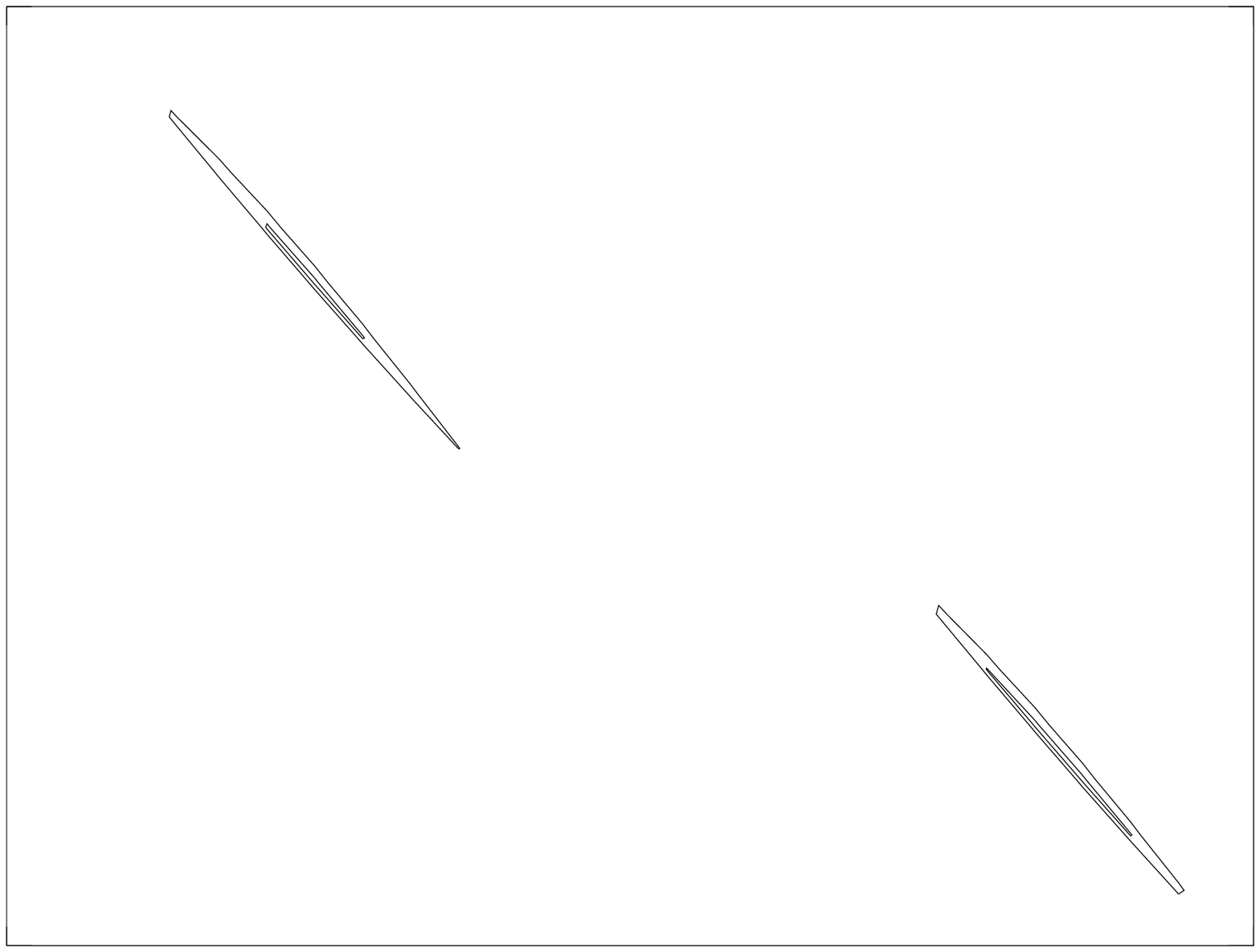}}
\end{picture}
\end{center}
\caption{The 68\%, 90\% and 99\% contours for a linear least
squares fit of R93, R96d, R98 X00a, X00b and CH00 (continuous elliptical
regions). Parallel lines are the 68\%
and 90\% contours of X00a PN; tiny elliptical regions are the 68\% and
90\% contours for the combined R93 and R96 datasets (see zoom). }    
\label{stage2}
\end{figure} 

We first performed an MLP assuming $\dot P=0$ on each of the longer
pointing: R93, R96d, X00a, X00b (figure 1). There is no
ambiguity in the period determinations and a linear least square fit
using the 68\% formal errors in the 
MLP gives $P_0 = 8.39113\pm0.00011$~s, $\dot
P=0.0\pm5.5\times10^{-13}$ s/s (here and in the following $P_0$ is
referenced
to the start of the R93 run). This upper limit on $\dot P$ permits an
unambiguous determination of the peaks in the Ch00 and R98 power 
spectra. Adding these to the linear square fit gives $P_0 = 8.39107 \pm
0.00005$ and $\dot P =
2.7\times10^{-13}\pm2.5\times10^{-13}$. The 68, 90 and 99\% confidence
intervals are shown in figure \ref{stage2}, as well the 68 and 90\%
intervals derived from X00a. With the improved $(P_0 \, ,\dot P)$ values,
we performed an MLP on the combined R93 and R96 datasets. As a result, the
confidence contour break up into small region (aliases) in the  $(P_0 \,
,\dot P)$ plane (see zoom in figure 2). With this further restriction, we
finally do the MLP on all data. 

We derive two pairs of values $(P_0 \, ,\dot P)$ which cannot be further
discriminated between on statistical grounds (table 
\ref{szane-D1_2_tab:tab2}). 

\begin{table}[bht]
  \caption{The two pairs of best fitting $(P_0 \, ,\dot P)$ values. 
$\Delta \chi^2$ is the difference between the $\chi^2$ of solution (2) and 
that of solution (1). Confidence levels can be read from figure
\ref{szane-D1_2_fig:fig3}.}
  \label{szane-D1_2_tab:tab2}
  \begin{center}
    \leavevmode
    \footnotesize
    \begin{tabular}[h]{llll}
      \hline \\[-5pt]
Label & $P_0 (s) $   & $\dot P (s/s) $       & $\Delta \chi^2$ \\[+5pt]
      \hline \\[-5pt]
(1) & $8.39109273$ & $5.409\times10^{-14}$ &    \\
(2) & $8.39109148$ & $3.749\times10^{-14}$ & 1.3  \\
      \hline \\
      \end{tabular}
  \end{center}
\end{table}
\begin{figure}[bht]
  \centerline{
    \epsfig{file=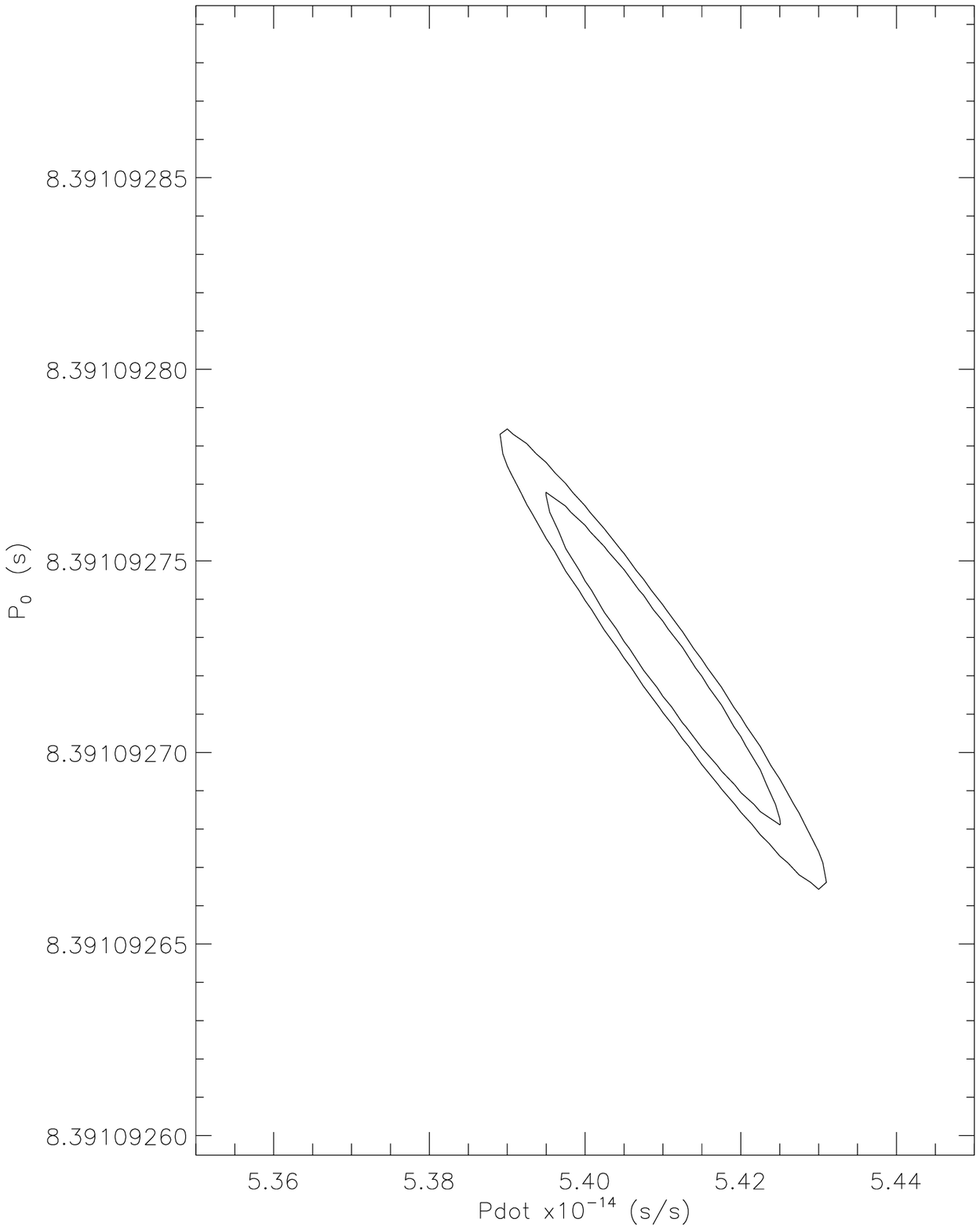, width=4.5cm, height=4.cm}
    \epsfig{file=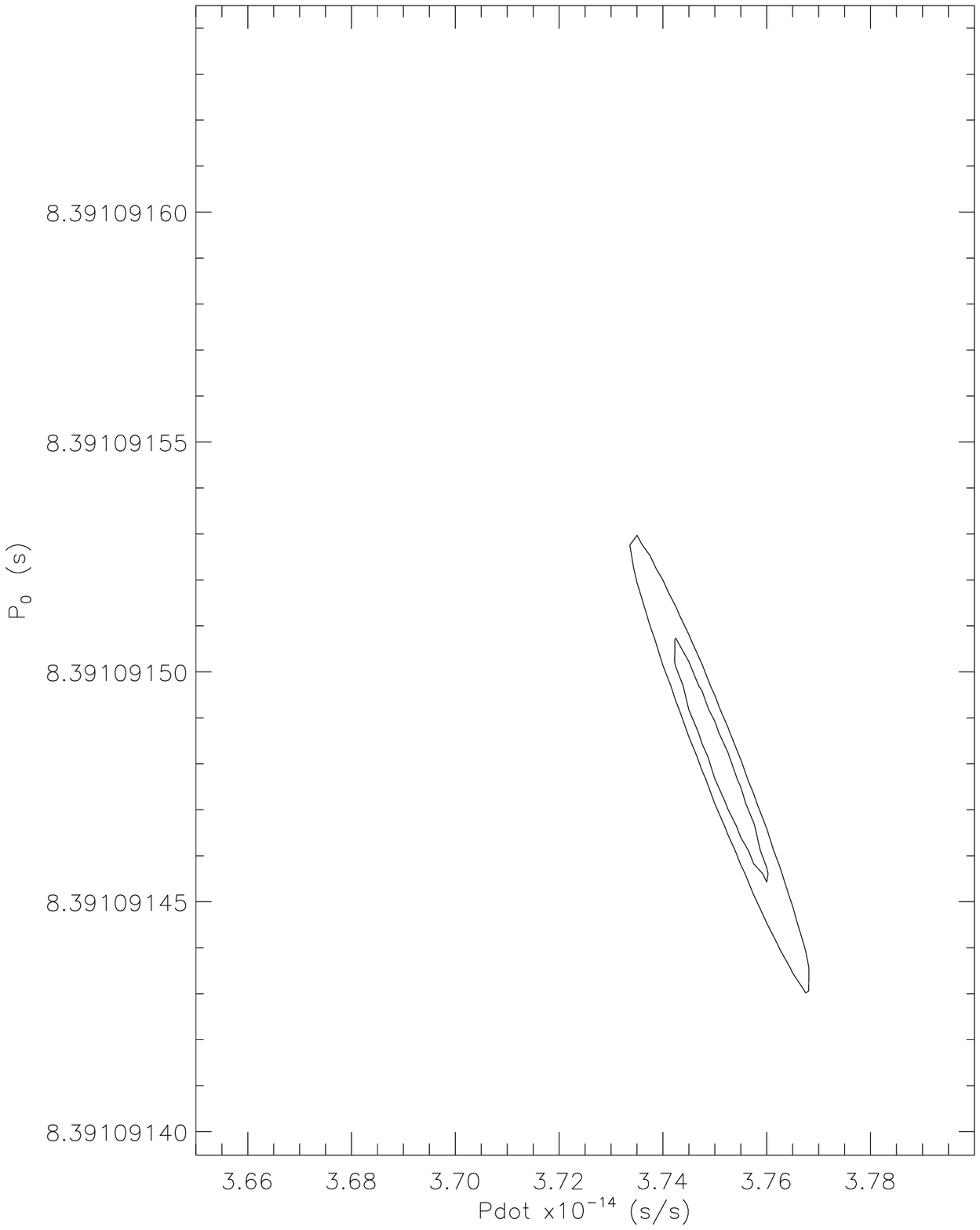, width=4.5cm, height=4.cm}}
\caption{The 68\% and 90\% MLP contours for $P_{0}$ and $\dot P$ to the
complete dataset except for the R98 data, for the two solutions 
(1) and (2).}  
\label{szane-D1_2_fig:fig3}
\end{figure}

\begin{figure}[bht]
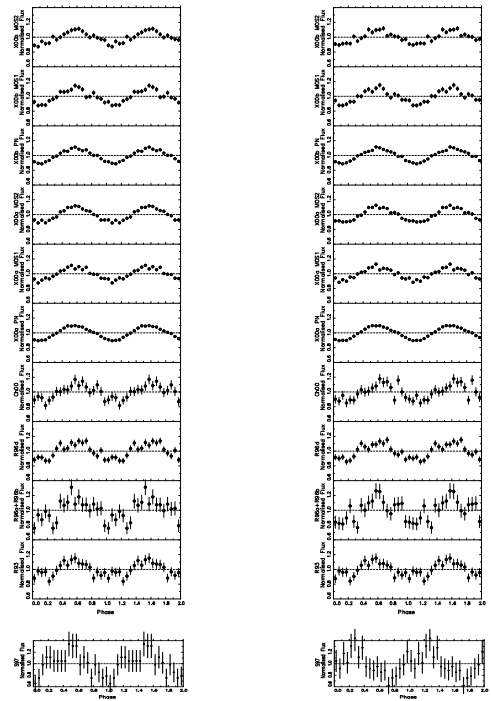

\begin{center}
\setlength{\unitlength}{1cm}
\begin{picture}(18,9)
\put(0.,1){\includegraphics{szane-D1_2_fig4a.ps}}
\put( 4.0,1){\includegraphics{szane-D1_2_fig4b.ps}}
\put( 1.65,-1.0){\includegraphics{szane-D1_2_fig4c.ps}}
\put( 5.65,-1.0){\includegraphics{szane-D1_2_fig4d.ps}}
\end{picture}
\end{center}
\caption{The datasets folded on the $(P_0 \, ,\dot P)$ solution (1) (left)
and (2)
(right).}
\label{szane-D1_2_fig:fig4}
\end{figure}

We have folded the data on both $(P_0 \, ,\dot P)$ solutions (1) and (2)
in table \ref{szane-D1_2_tab:tab2} to check the relative phasing of
all individual runs (figure \ref{szane-D1_2_fig:fig4}). We have then
performed the {\it a posteriori} check
with the {\sl BeppoSax} data, which phase on correctly
with solution (1), but not (2) (see the lowest panel of figure
\ref{szane-D1_2_fig:fig4}). This suggest to select solution (1) as the
most likely timing parameters. In any case, for the purposes of our
further discussion, the difference between the $\dot P$ in (1) and (2) is
not significant: both acceptable fits to the data have
$3\times10^{-14} < \dot P < 6\times 10^{-14}$. This is the most accurate
spin-down measure presented so far for a dim NS and, for the first time,
it allows a discrimination between the proposed models. 

\section{Discussion}
\label{szane-D1_2_sec:disc}

The refined value of $\dot P$ reported here is consistent with,
but two orders of magnitude lower than the extrema of the range reported
by \cite*{szane-D1_2:ha97}. The first implication is that
\object{RX~J0720.4-3125} is unlikely to
be spinning down under propeller torques (\cite{szane-D1_2:al01}). In
this case, under the assumption that the X-ray luminosity of the source 
($L_{\rm X} \approx 2 \times
10^{31} d_{100}^2$ erg~cm$^{-2}$~s$^{-1}$, where 
$d_{100} = d/100 \, {\rm pc}$ and $d$ is the
distance,
\cite{szane-D1_2:ha97}) is
supplied by energy dissipation, it should be: 
\begin{equation}
2 \times 10^{-11} d_{100}^2  \leq \dot P \leq 2 \times
10^{-9} d_{100}^2  \   { {\rm s } \over {\rm s} } \, .
\end{equation}    

The scenario is still consistent with the spin-down recently 
measured for \object{RBS~1223} (\cite{szane-D1_2:ham01}), but the value of
$\dot P$ reported here for \object{RX~J0720.4-3125} is well below this
range. 
That also make less plausible
an interpretation of the hardness ratio profile in terms of 
absorbing matter, co-rotating in the magnetosphere
(\cite{szane-D1_2:cr01}). 
The observed behaviour is more probably explained by the
angle-dependent properties of the emitted radiation.     

On the other hand, the slow-down rate of \object{RX~J0720.4-3125} is still
considerable. The other plausible mechanism which may account for such
large and stable value of $\dot P$ is magnetic breaking.
For a dipolar magnetic field $\dot P \approx 10^{-15} \left ( B / 
10^{12} G  \right )^2/P$~s/s, which gives $B = 2.13
\times 10^{13}$~G~\footnote{Here and in the following we
specify the discussion to solution (1) of table
\ref{szane-D1_2_tab:tab2}}. A scenario in which this source is
powered by accretion from the interstellar medium must be therefore ruled
out: for the present values of $P$ and $B$ the co-rotating magnetosphere
will prevent the incoming material to penetrate below the Alfven radius.

The corresponding spin-down age is $t_{\rm sd} =  \dot P / \left ( 2P
\right      
) \sim 2.48 \times 10^6  \, {\rm yr}$, 
which, given the numerous uncertainties, is marginally compatible with
that inferred by the cooling curves (a few $10^5$ yrs for a surface
temperature of $\sim 80$~eV, e.g. 
\cite{szane-D1_2:khy01}, \cite{szane-D1_2:kyg01}, \cite{szane-D1_2:sc97},
\cite{szane-D1_2:sc99}). The
discrepancy is
less significant if we notice that what we are probably observing in the
X-ray is a region of limited size which
is kept hotter than the average star surface, as inferred by the analysis
of the pulse-shape (\cite{szane-D1_2:cr01}).

On the other hand, $t_{\rm sd}$ is representative of the true age of the
source only in the case in which the magnetic field remained almost
constant during the star evolution. The same condition applies for the
validity of the cooling curves mentioned above, which do not include the
extra input of energy released in the neutron star in case of field
decay. It is therefore of fundamental importance to address the field   
evolution. There are three
mechanisms which are typically proposed for inducing field-decay:
ambipolar diffusion in the solenoidal or irrotational mode and Hall
cascade (\cite{szane-D1_2:gr92}, \cite{szane-D1_2:hk98}, 
\cite{szane-D1_2:cgp00}). By using the simplified expressions of 
\cite*{szane-D1_2:cgp00} for the decay laws, we have estimated the source
age and the value of the magnetic field at the birth of the neutron star,
$B_0$ (table \ref{szane-D1_2_tab:tab3}). 

\begin{table}
\begin{center}
\begin{tabular}{lll}
\hline
B-Decay Mechanism & $B_0$ & age \\
      &  $10^{13}$ G &  (years) \\
\hline
Hall Cascade & 119.2 & $4.5 \times 10^4$\\
Ambipolar diffusion, irrotational mode & 1.9 & $3.3 \times 10^6$\\
Ambipolar diffusion, solenoidal mode & 4.1 & $1.6 \times 10^6$\\
\hline
\end{tabular}   
\caption{Predicted source age and primordial field for three different 
mechanisms of decay, simulated as in Colpi et~al. 2000. The present values
of $P$ and $\dot P$ are those of solution (1) in table 
\ref{szane-D1_2_tab:tab2}. In all cases, the source is assumed to be born
with $P = 1$~ms.}
\label{szane-D1_2_tab:tab3}
\end{center}
\end{table}    

As we can see, allowing for a mechanism involving
very fast decay, such as the Hall cascade, we find that the source is now
$\sim 4 \times 10^4$~yr old, and is born with a superstrong field $B_0
\sim 10^{15}$~G. Such a young age is only marginally compatible with the
absence of a remnant and, more important, is not compatible with the
low observed X-ray luminosity. 
Underluminous models have been 
presented by \cite*{szane-D1_2:khy01}, who accounted for the
enhanced neutrino cooling in presence of strong neutron superfluidity.
These
solutions may match an age of $\sim 10^4$~yrs for
\object{RX~J0720.4-3125}, but, as
discussed by the same authors, they must probably be rejected since they
fail in
the comparison with observational data of a sample of other neutron stars. 

On the other hand, both mechanisms involving
ambipolar diffusion predict a magnetic field quite stable over the
source lifetime and close to the actual value. Accordingly, the
predicted age is $\sim 10^6$ years in all cases, close to $t_{\rm sd}$.
Below $\sim 10^{14}$~G the cooling curves are not significantly
influenced by decay through ambipolar diffusion (\cite{szane-D1_2:hk98}),
thus, as in the case of constant $B$ discussed above, the scenario is
compatible with the observed luminosity. The larger age is also compatible
with the absence of a remnant.

If our conclusions are valid, the connection between dim INS and AXPs is
not so obvious. \object{RX~J0720.4-3125} has a strong, but not
superstrong, field
which is compatible with those of the canonical radio-pulsars which have
passed the death line. On the other hand, having excluded accretion,
what mechanism causes an X-ray emission concentrated in a fraction of
$\sim 60$\% of the star surface remain a mystery, as well the related
question about the validity of using the observed blackbody temperature to
locate the source in the cooling diagram. The variation of the surface
cooling temperature with the latitude predicted so far for strong fields  
(\cite{szane-D1_2:gh83}, \cite{szane-D1_2:po96})
is too smooth to explain the (relatively) small size of the emitting
region, and different explanations are required.

\begin{acknowledgements}

We are grateful to Mat Page for his advice on the maximum likelihood
methods,
and to Darragh O'Donoghue for pointing us to the Raleigh transform, and
for the use of his Eagle Fourier transform code used in the preliminary
stages of this work. We are grateful to Monica Colpi for lots of useful
discussions and to Ulrich Geppert for providing the cooling curves
computed allowing for the extra-heating due to $B$-decay from Hall
cascade. 

\end{acknowledgements}

\end{document}